\documentclass[11pt,oneside, a4paper]{article}

\ifx\pdfoutput\undefined
\usepackage[dvips,bookmarks=false]{hyperref}
\else
\usepackage{hyperref}
\fi
\hypersetup{colorlinks,bookmarksopen,bookmarksnumbered,citecolor=blue,
linkcolor=black,pdfstartview=FitH,urlcolor=blue}

\usepackage{graphicx}
\usepackage{amssymb}
\usepackage{cite}
\usepackage{bm}
\usepackage{indentfirst}
\usepackage{amsmath}
\usepackage{hhline}
\usepackage{multirow}

\def\be{\begin{equation}}
\def\ee{\end{equation}}
\newcommand{\bea}{\begin{eqnarray}}
\newcommand{\eea}{\end{eqnarray}}

\begin{document}
\begin{titlepage}

\begin{flushright}
LPT-ORSAY-16-61
\end{flushright}

\begin{center}

\vspace{1cm}
{\large\bf 
Electric dipole moments of charged leptons with sterile fermions
}
\vspace{1cm}

\renewcommand{\thefootnote}{\fnsymbol{footnote}}
Asmaa Abada$^1$\footnote[1]{asmaa.abada@th.u-psud.fr}
and 
Takashi Toma$^1$\footnote[2]{takashi.toma@th.u-psud.fr}
\vspace{5mm}

{\it
$^{1}$Laboratoire de Physique Th\'eorique, CNRS, \\
Univ. Paris-Sud, Universit\'e Paris-Saclay, 91405 Orsay, France
\vspace*{.2cm} 
}

\vspace{8mm}

\abstract{
We address the impact of sterile fermions on charged lepton electric
dipole moments. 
We show that in order to have a 
non-vanishing contribution to electric dipole moments, the minimal
 extension
necessitates the addition of at least two sterile fermion
 states. Sterile
neutrinos can give significant contributions to the charged lepton
electric dipole moments if the masses of the non-degenerate sterile
states are both above the electroweak scale. In addition, the Majorana
nature of neutrinos is also important.
Furthermore, we apply the computations of the electric dipole moments
 for the most minimal
realisation of the Inverse Seesaw mechanism, in which the Standard Model
is extended by two right-handed neutrinos and two sterile fermion
states. We show that the two pairs of (heavy) pseudo-Dirac mass
eigenstates can give significant contributions to the electron electric
dipole moment, lying close to future experimental sensitivity. 
We further discuss the possibility of beyond the minimal Inverse Seesaw
models and of having a successful leptogenesis in this framework. 
}

\end{center}
\end{titlepage}

\renewcommand{\thefootnote}{\arabic{footnote}}
\setcounter{footnote}{0}

\setcounter{page}{2}


\section{Introduction}
Electric Dipole Moments (EDMs) are CP violating observables which are 
sensitive to new physics. 
In the Standard Model (SM), the Cabibbo-Kobayashi-Maskawa matrix
is the only source of CP violation, and the charged lepton EDMs
are induced at four-loop level. The corresponding order of magnitude is roughly given
by 
\begin{equation}
|d_e|/e\sim\frac{\alpha_W^3\alpha_sm_e}{246(4\pi)^4m_W^2}J_{CP}\sim
10^{-38}~\mathrm{cm},
\label{eq:edm_sm}
\end{equation}
where $J_{CP}\equiv\mathrm{Im}\left(V_{us}V_{cs}^*V_{cb}V_{ub}^*\right)$
is the Jarlskog invariant whose value is fixed to be
$J_{CP}\approx10^{-5}$ by the experiments~\cite{Patrignani:2016}. 
For the other charged leptons, muon and tau, the EDMs in the SM are estimated by replacing the electron mass
$m_e$ in Eq.~(\ref{eq:edm_sm}) by the mass of muon or tau. 
On the other hand, the current bounds of the charged lepton EDMs are
given by 
\begin{eqnarray}
|d_e|/e
\hspace{-0.2cm}&<&\hspace{-0.2cm}
8.7\times10^{-29}~\mathrm{cm}\hspace{1cm}\mbox{(ACME Collaboration)~\cite{Baron:2013eja}},\\
|d_\mu|/e
\hspace{-0.2cm}&<&\hspace{-0.2cm}
1.9\times10^{-19}~\mathrm{cm}\hspace{1cm}\mbox{(Muon $g-2$ Collaboration)~\cite{Bennett:2008dy}},\\
|d_\tau|/e
\hspace{-0.2cm}&<&\hspace{-0.2cm}
4.5\times10^{-17}~\mathrm{cm}\hspace{1cm}\mbox{(Belle Collaboration)~\cite{Inami:2002ah}}.
\end{eqnarray}
Thus the SM predictions of the charged lepton EDMs are
far below the current experimental bounds. 
However if one considers some extensions of the SM, new contributions
may be able to generate larger values close to the current
experimental bounds or future sensitivities. 

There are a number of extensions of the SM to address some of its
observational caveats such as non-zero neutrino masses, existence of dark matter
and baryon asymmetry in the Universe.
Adding sterile fermions to the SM is one of the most economical extensions of the SM. 
We consider models extended with only sterile fermions,
and compute the charged lepton EDMs. 
First, we do not impose a specific seesaw mechanism for neutrino masses like
Type-I Seesaw, Inverse Seesaw or Linear Seesaw mechanism (the effective model). 
Second, the minimal Inverse Seesaw model is considered as a more
specific case.

\section{The Effective Model}

The SM is extended with $N$ sterile fermions. 
The relevant interactions in the Feynman-'t Hooft gauge are given by
\begin{equation}
\mathcal{L}=
-\frac{g_2}{\sqrt{2}}U_{\alpha
i}W_{\mu}^-\overline{\ell_\alpha}\gamma^{\mu}P_L\nu_i
-\frac{g_2}{\sqrt{2}}H^-\overline{\ell_\alpha}
\left(\frac{m_\alpha}{m_W}P_L-\frac{m_i}{m_W}P_R\right)\nu_i
+\mathrm{H.c.},
\label{eq:int}
\end{equation}
where $H^-$ is the Goldstone boson absorbed by the $W$ gauge boson, 
$U_{\alpha i}$ is the mixing matrix of the neutrinos
($\nu_\alpha=U_{\alpha i}\nu_i$),
$\ell_\alpha$ is the charged lepton and $\nu_i$ is the mass eigenstates
of the neutrinos. The indices denote $\alpha=e,\mu,\tau$ and $i,j=1,\cdots,N$.

We do not fix any neutrino mass generation mechanism and the neutrino
masses $m_i$, and thus the mixing matrix elements $U_{\alpha i}$ are regarded
as independent parameters. 
The mixing matrix $U_{\alpha i}$ in Eq.~(\ref{eq:int}) is the sub-matrix
of the whole mixing matrix $U_{ij}$ which is appropriately
parametrized.

\begin{figure}[t]
\begin{center}
\includegraphics[scale=0.55]{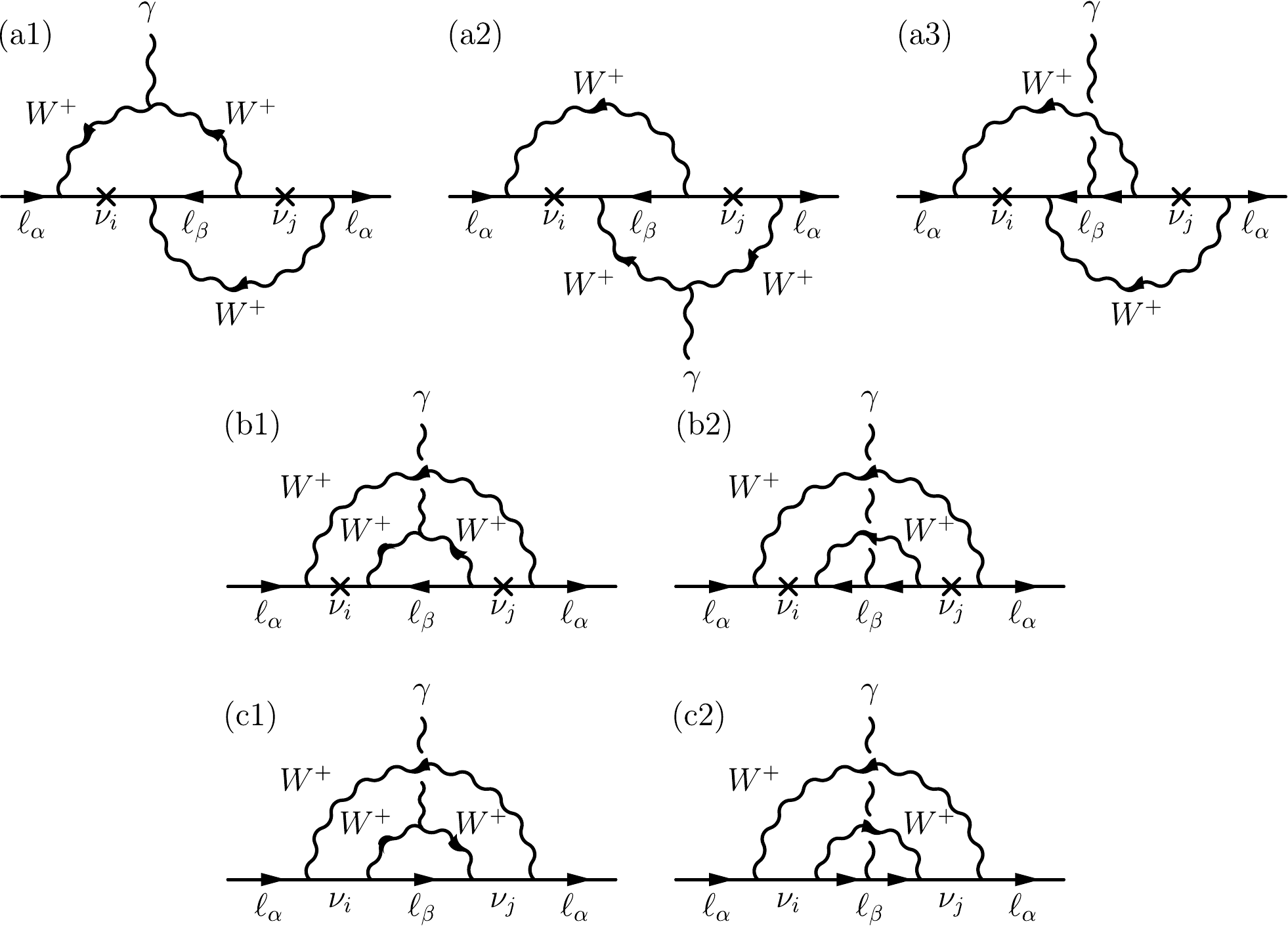}
\caption{Diagrams contributing to charged lepton EDMs.}
\label{fig:diagram1}
\end{center} 
\end{figure}

The leading contribution to the charged lepton EDMs is induced at two-loop
level via the diagrams shown in Fig.~\ref{fig:diagram1}, which includes
a total of $44$ diagrams in the Feynman-'t Hooft gauge. 
We compute all the diagrams using FeynCalc~\cite{Mertig:1990an} in order to obtain
an analytic result which is given in the Appendix of the reference~\cite{Abada:2015trh}. 
The charged lepton EDMs are formally written as
\begin{equation}
d_\alpha=-\frac{g_2^4em_\alpha}{4(4\pi)^4m_W^2}
\sum_{\beta}\sum_{i,j}\sqrt{x_ix_j}
\Bigl[
J_{ij\alpha\beta}^MI_M(x_i,x_j)
+J_{ij\alpha\beta}^DI_D(x_i,x_j)
\Bigr],
\label{eq:edm}
\end{equation}
where $x_i\equiv m_i^2/m_W^2$, $J_{ij\alpha\beta}^M$
and $J_{ij\alpha\beta}^D$ are the phase factors of the Majorana type and Dirac type
contributions defined by $J_{ij\alpha\beta}^M\equiv\mathrm{Im}\left(U_{\alpha
j}U_{\beta j}U_{\beta i}^*U_{\alpha i}^*\right)$ and 
$J_{ij\alpha\beta}^D\equiv\mathrm{Im}\left(U_{\alpha
j}U_{\beta j}^*U_{\beta i}U_{\alpha i}^*\right)$, and the charged lepton
masses can be safely neglected compared to the $W$ gauge boson mass $m_\alpha^2\ll m_W^2$.
From the definition of the phase factors $J_{ij\alpha\beta}^M$ and
$J_{ij\alpha\beta}^D$, one can see that they are anti-symmetric under
the exchange $i\leftrightarrow j$. As a result, one can verify that only
the anti-symmetric part of
the loop functions $I_M$ and $I_D$ can contribute to the EDMs in Eq.~(\ref{eq:edm}). 

In the case of $N=1$, the EDM expression can be simplified as~\cite{Abada:2015trh}
\begin{equation}
d_\alpha\approx
-\frac{g_2^4em_\alpha}{2(4\pi)^4m_W^2}
\sum_{\beta}\sum_{i=1}^3\sqrt{x_ix_4}
J_{i4\alpha\beta}^DI_D(0,x_4). 
\end{equation}
Therefore, taking $I_D(0,x_4)\sim1$ and $x_i\sim10^{-24}~(i=1,2,3)$, 
the predicted electron EDM is evaluated to be
$|d_e|/e\lesssim10^{-39}~\mathrm{cm}$, which is too small compared to the
current experimental bound. 
For $N=2$, the EDM expression can be written as 
\begin{equation}
d_\alpha\approx
-\frac{g_2^4em_\alpha}{2(4\pi)^4m_W^2}\sqrt{x_4x_5}
\Bigl[
J_\alpha^MI_M(x_4,x_5)+
J_\alpha^DI_D(x_4,x_5)
\Bigr],
\end{equation}
where $J_\alpha^{M/D}=\sum_\beta J_{45\alpha\beta}^{M/D}$. 
Only heavy sterile states in the loop provide a dominant contribution, and
the EDMs can be potentially large enough to be detected. 
Two different sterile states enter in the loops as shown in Fig.~\ref{fig:diagram1}. 
Thus one can see that at least two sterile fermions are needed to induce sizeable EDMs. 
If $N$ sterile fermions are added to the SM, the predicted EDMs are
expected to increase with the factor $N(N-1)/2$. 

\begin{figure}[t]
\begin{center}
\includegraphics[scale=0.55]{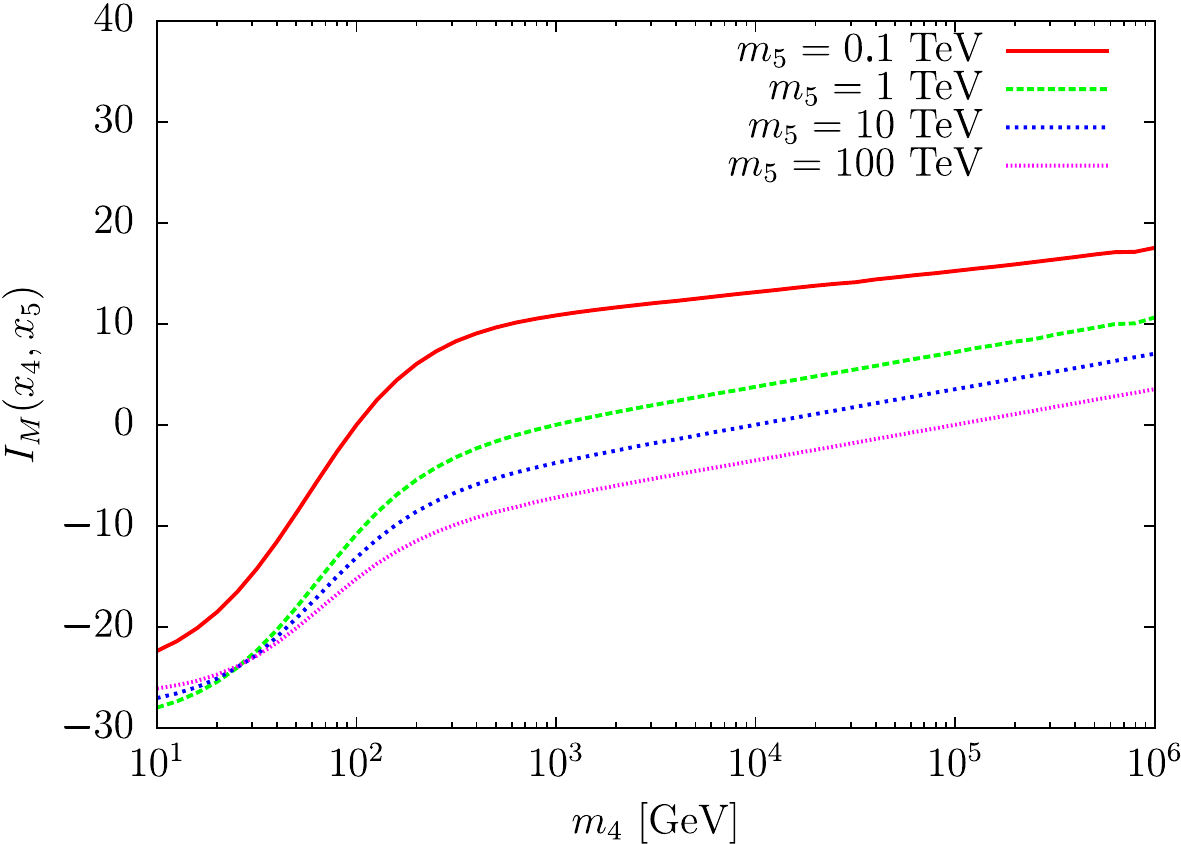}
\includegraphics[scale=0.55]{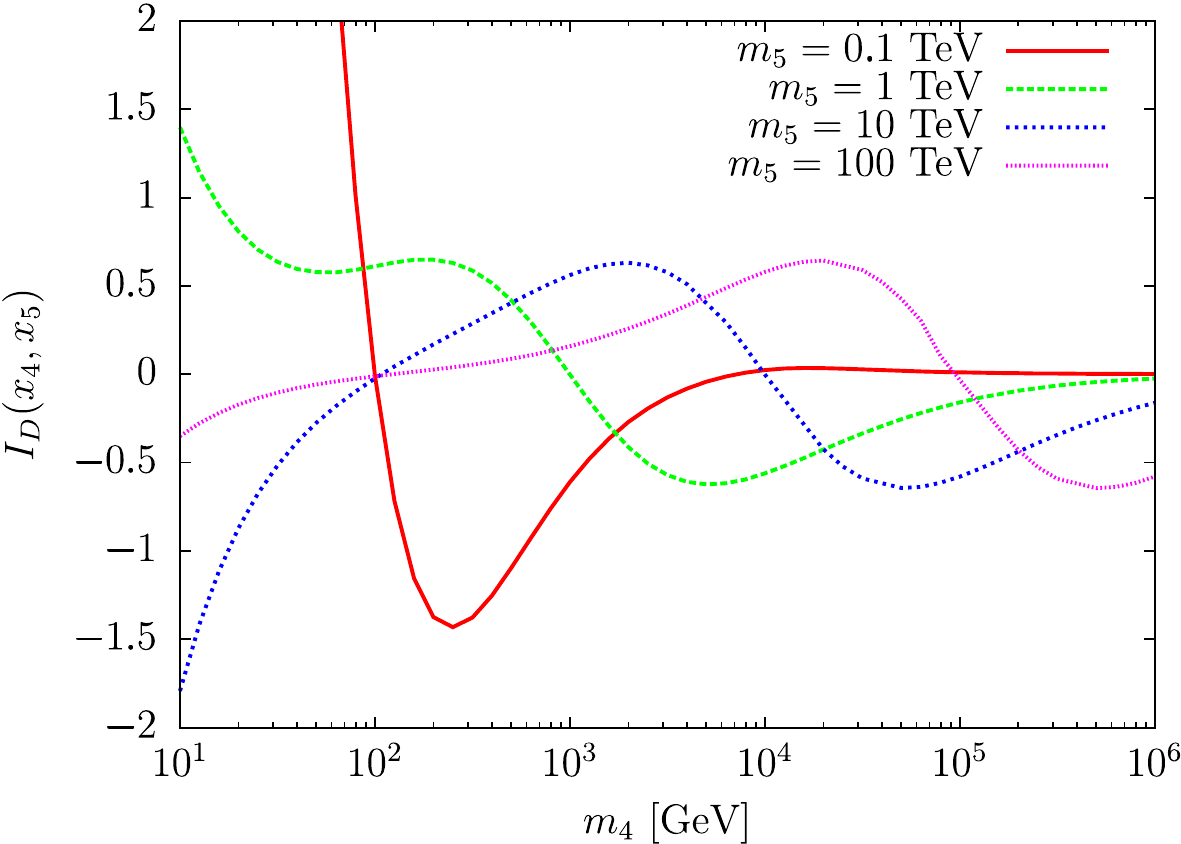}
\caption{Loop functions $I_D$ and $I_M$ as a function of $m_4$ for fixed
 $m_5$.}
\label{fig:loop}
\end{center} 
\end{figure}

The behavior of the loop functions is shown in
Fig.~\ref{fig:loop}, where several values for $m_5$ are chosen. 
From the plots, one can see that $I_M\gg I_D$ for $x_4,x_5\gg1$. 
Although the loop function $I_D$ can be larger than $I_M$ if
$x_4,x_5\ll1$, the predicted EDMs are very suppressed since
the loop functions can be expanded in $x_4$ and $x_5$ as
$I_{M/D}(x_4,x_5)\propto(x_4-x_5)\ll1$, due to the anti-symmetry. 

\begin{figure}[t]
\begin{center}
\includegraphics[scale=0.5]{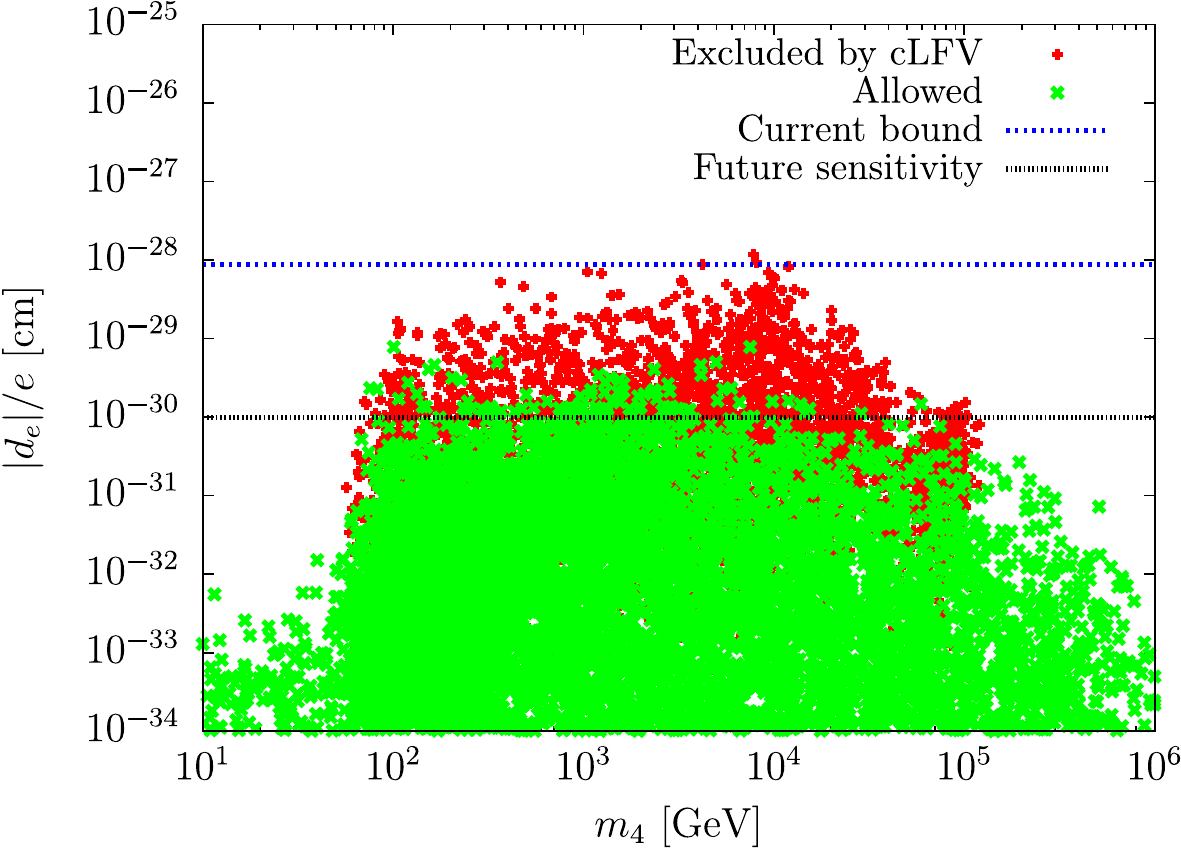}
\includegraphics[scale=0.5]{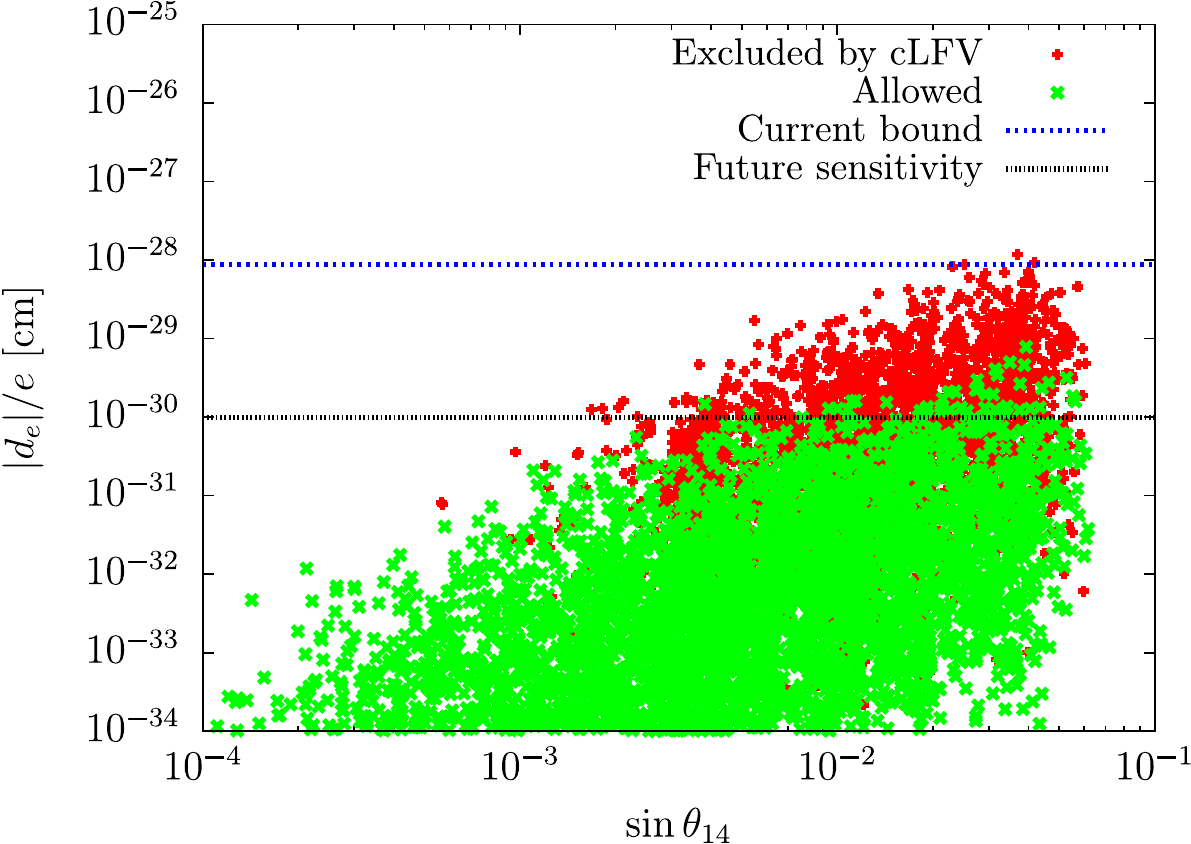}
\caption{Predicted electron EDM with the experimental and theoretical constraints. 
The red points are excluded by the charged lepton flavour
 violation bounds while the green points are allowed by all the constraints.}
\label{fig:edm1}
\end{center} 
\end{figure}

\begin{figure}[t]
\begin{center}
\includegraphics[scale=0.55]{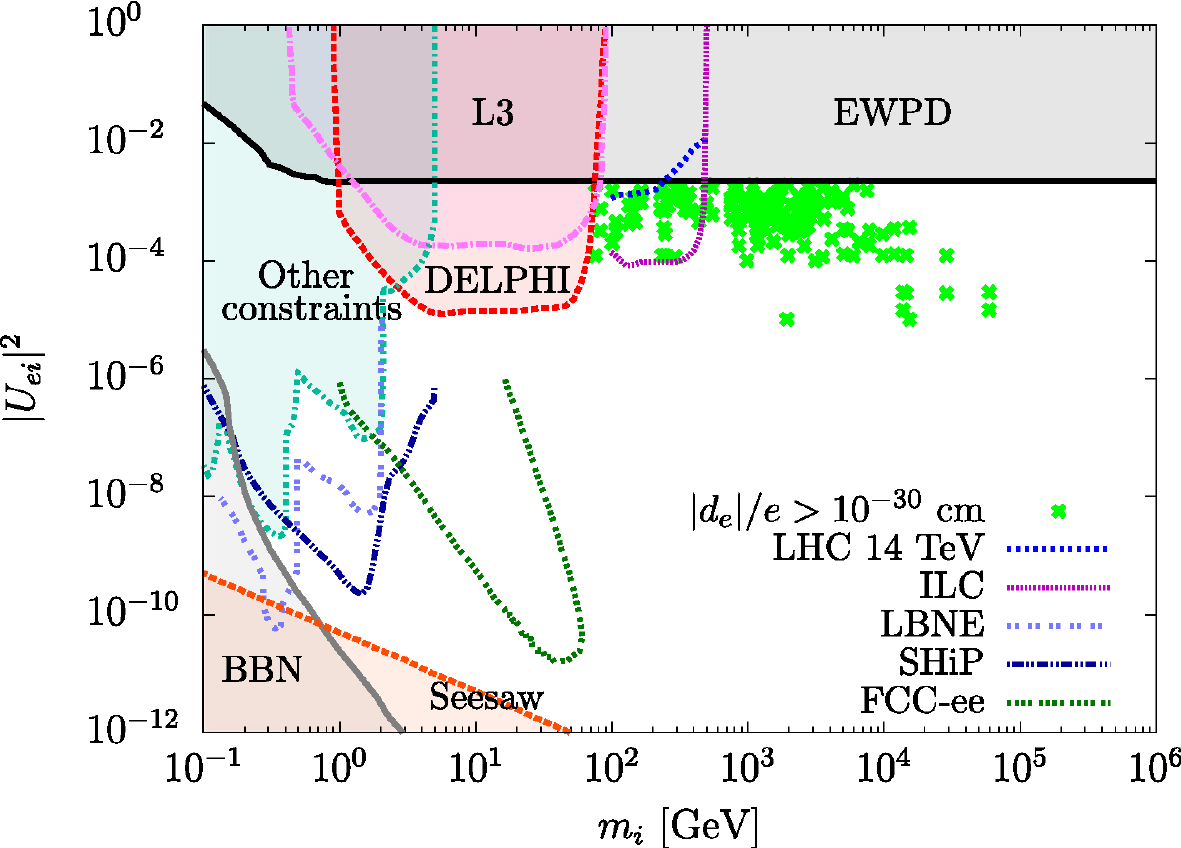}
\caption{Allowed parameter space in the ($m_i$-$|U_{ei}|^2$) plane. The
 colored regions are excluded. The green points can induce an electron
 EDM larger than the future sensitivity $|d_e|/e=10^{-30}~\mathrm{cm}$.}
\label{fig:edm2}
\end{center} 
\end{figure}

The experimental and theoretical constraints are taken into account:
these include neutrino oscillation data, charged lepton flavour
violation, direct production of the sterile neutrinos at colliders,
electroweak precision tests including $W$ boson decay, $Z$ invisible
decay, lepton flavour universality of meson decays, non-unitarity of the
mixing matrix $U_{\alpha i}$ and perturbative unitarity
bound~\cite{Deppisch:2015qwa}. 
With these constraints, the electron EDM is computed as shown in
Fig.~\ref{fig:edm1}. 
The red points are excluded by the charged lepton flavour violation bounds, while
the green points are allowed by all the above experimental and theoretical
constraints. The blue dotted line is the current bound of the electron EDM 
as measured by the ACME collaboration, and the black dotted line shows the
future sensitivity of the upgraded ACME collaboration. 
The region of $m_4\lesssim\mathcal{O}(100)~\mathrm{GeV}$ is excluded by
the electroweak precision tests. 
The upper region is mainly constrained by the charged lepton flavour
violation bounds and the upper right region is excluded by perturbative
unitarity bound. 
All the green points are below the current bound of the electron EDM, however
some points can be within reach of the future sensitivity
$|d_e|/e=10^{-30}~\mathrm{cm}$. 
Figure~\ref{fig:edm2} shows the allowed parameter space from all the
constraints in the ($m_i$-$|U_{ei}|^2$) plane. 
The colored regions are excluded. 
The green points represent the parameter space inducing an electron EDM larger than
the future sensitivity $|d_e|/e\gtrsim10^{-30}~\mathrm{cm}$. 
One can see that some green points can also be testable by a future
ILC experiment (violet line).
The EDMs for the other charged leptons (muon and tau) are also computed and compared to
the current bounds and future sensitivities. 
However the order of the magnitude for the predictions is too small to be
detected in near future. 

\section{The Inverse Seesaw Models}

We consider the Inverse Seesaw as a specific model. 
The SM is extended with both right-handed neutrinos $N^c$ and singlet
fermions $s$. 
At least two pairs of these are required to accommodate neutrino oscillation data. 
This minimal Inverse Seesaw model is denoted as $(2,2)$-model~\cite{Abada:2014vea}. 
In the minimal Inverse Seesaw model, the neutrino mass matrix is given by
\begin{equation}
M=\left(
\begin{array}{ccccccc}
0 & 0 & 0 & d_{11} & d_{12} & 0 & 0\\
0 & 0 & 0 & d_{21} & d_{22} & 0 & 0\\
0 & 0 & 0 & d_{31} & d_{32} & 0 & 0\\
d_{11} & d_{21} & d_{31} & m_{11} & m_{12} & n_{11} & n_{12}\\
d_{12} & d_{22} & d_{32} & m_{21} & m_{22} & n_{12} & n_{22}\\
0 & 0 & 0 & n_{11} & n_{12} & \mu_{11} & \mu_{12}\\
0 & 0 & 0 & n_{12} & n_{22} & \mu_{12} & \mu_{22}\\
\end{array}
\right),
\end{equation}
in the basis $(\nu_{e},\nu_{\mu},\nu_{\tau},N_1^c,N_2^c,s_1,s_2)^T$. 
The matrix element $d_{ij}$ is the Dirac mass term between the
left-handed and the right-handed neutrinos given by electroweak symmetry
breaking, and $n_{ij}$ is the Dirac mass term between the right-handed
neutrinos $N^c$ and the singlet fermions $s$. 
The matrix elements $m_{ij}$ and $\mu_{ij}$ are Majorana mass terms violating lepton
number conservation. These Majorana mass terms are (naturally) assumed to
be much smaller than the other Dirac mass terms. 
The mass matrix $M$ can be diagonalized as
$U^TMU=\mathrm{diag}(m_1,m_2,\cdots,m_7)$ with the mass hierarchy $|\mu|,|m|\ll|d|\ll|n|$. 
The mass eigenvalues $m_{1,2,3}$ correspond to the active light neutrino
masses. 
The heavy mostly sterile states turn out to be nearly degenerate as
$m_4\approx m_5$ and $m_6\approx 
m_7$ due to the nature of Inverse Seesaw models. 
This mass degeneracy means that the heavy sterile states construct pseudo-Dirac fermions. 

The relevant diagrams to the charged lepton EDMs in the minimal Inverse
Seesaw model are the same as those of the effective model shown
in Fig.~\ref{fig:diagram1}. 
However since the heavy sterile states are pseudo-Dirac fermions with
nearly degenerate
masses, the Majorana type contribution is negligible. 
Thus, one can focus on the Dirac type diagrams (c1)
and (c2) in Fig.~\ref{fig:diagram1}, and express the EDM formula as 
\begin{equation}
d_{\alpha}\approx
-\frac{g_2^4em_\alpha}{2(4\pi)^4m_W^2}J^DI_D^\prime(x_4,x_6), 
\end{equation}
where $I_D^\prime(x_4,x_6)$ is the reduced loop function whose analytic
formula and numerical evaluations are given in the reference~\cite{Abada:2016awd}.

\begin{figure}[t]
\begin{center}
\includegraphics[scale=0.55]{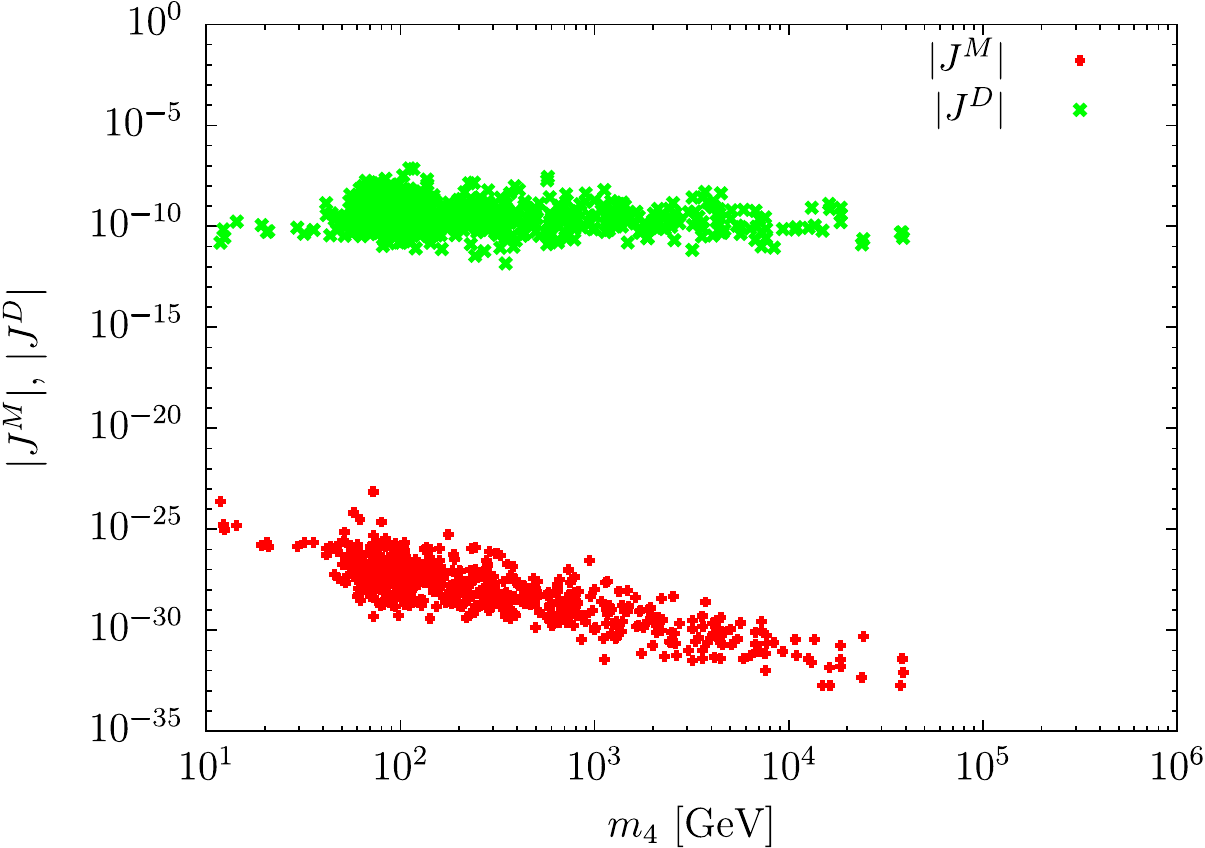}
\includegraphics[scale=0.55]{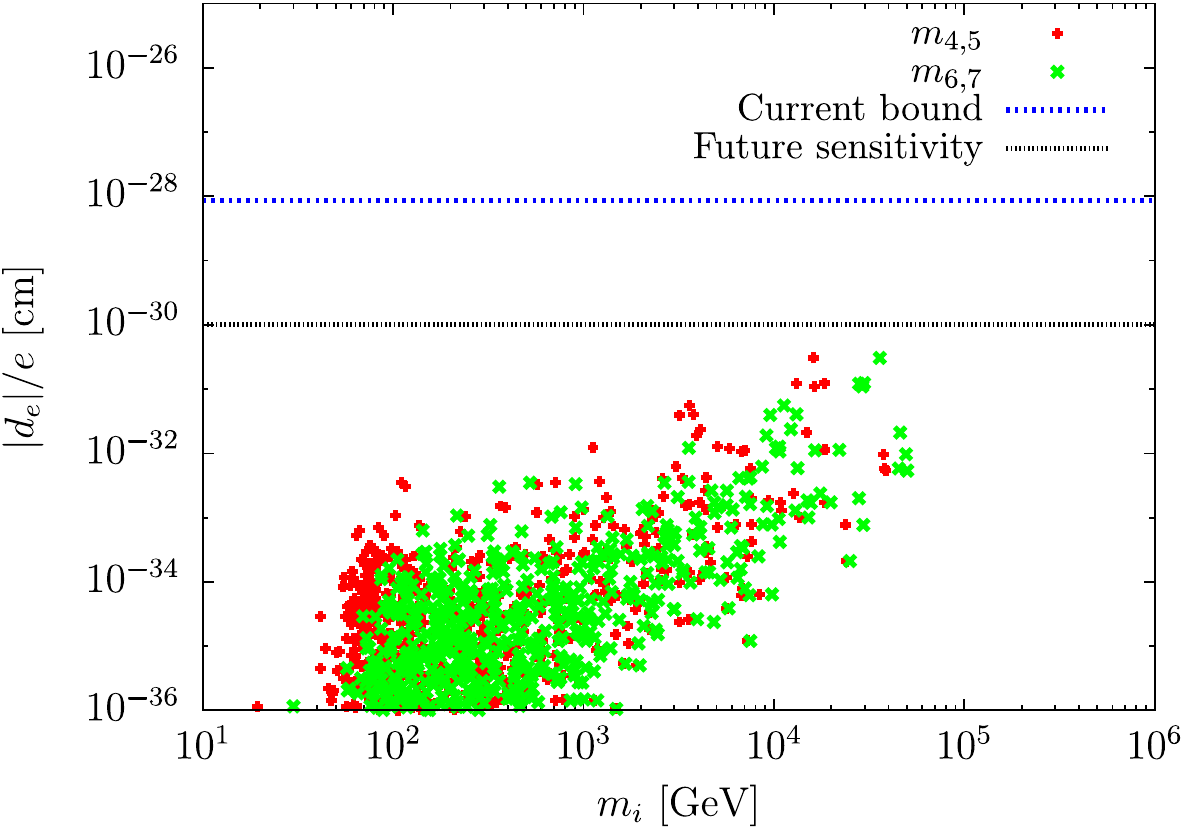}
\caption{Charged lepton EDMs in the $(2,2)$ Inverse Seesaw model.}
\label{fig:edm_inv}
\end{center} 
\end{figure}

The numerical computations for the phase factors $|J^M|$ and $|J^D|$, 
and the predicted electron EDM are shown in Fig.~\ref{fig:edm_inv} where all the
relevant constraints which have been considered in the effective model are
imposed. 
From the left plot of Fig.~\ref{fig:edm_inv}, one can see that $|J^D|$ is
much larger than $|J^M|$, as expected because of the pseudo-Dirac
heavy sterile states. 
The right plot shows that the maximum value of the predicted
electron EDM is slightly below the future prospects,
$|d_e|/e=10^{-30}~\mathrm{cm}$. 
Similar to the effective model, the other charged lepton (muon and tau)
EDMs have also been computed, but the predictions are very far from the
current bounds and future prospects. 

In the above, we considered the minimal Inverse Seesaw model with two
pairs of the right-handed neutrinos and singlet sterile fermions. 
If $N>2$ pairs are added to the SM, the electron EDM is
roughly given by
\begin{equation}
d_e^{(N,N)}\sim
\frac{g_2^4em_e}{2(4\pi)^4m_W^2}
\left|
4N(N-1)\mathrm{Im}(U^4)I_D^\prime
\right|. 
\label{eq:nn}
\end{equation}
Thus one can see that the electron EDM is enhanced if more sterile
fermions are introduced. 
However the constraints from charged lepton flavour violation also become
stronger at the same time. The most stringent process is $\mu\to
e\gamma$ whose branching ratio can be roughly given by
\begin{equation}
\mathrm{Br}(\mu\to e\gamma)\sim 
\frac{\sqrt{2}G_F^2m_{\mu}^5}{\Gamma_{\mu}}(2N)^2|U|^4\leq
4.2\times10^{-13}. 
\label{eq:meg}
\end{equation}
Combining Eq.~(\ref{eq:nn}) and (\ref{eq:meg}), the enhancement factor
compared to the minimal ($2,2$)-model is given by
\begin{equation}
\left|\frac{d_e^{(N,N)}}{d_e^{(2,2)}}\right|\lesssim
2\left(1-\frac{1}{N}\right).
\end{equation}
Therefore one can see that the maximum enhancement factor is $2$ from
this rough estimation.

We may also have a correlation with another CP violating observable. 
In particular, the baryon asymmetry of the Universe may be generated by
resonant leptogenesis since the heavy sterile states are naturally
nearly degenerate in Inverse Seesaw models. 
In order to generate a sufficient baryon asymmetry of the Universe, the
out-of-equilibrium condition $\Gamma_{\nu_4}<H(T)|_{T=m_4}$ should be
satisfied for the lightest heavy sterile state, where $H(T)$ is the
Hubble parameter and the decay rate $\Gamma_{\nu_4}$ is given by 
\begin{equation}
\Gamma_{\nu_4}=\frac{g_2^2m_4^3}{16\pi m_W^2}\sum_{\alpha}|U_{\alpha
 4}|^2. 
\label{eq:out}
\end{equation}
Thus, the following condition can be derived
\begin{equation}
\sum_{\alpha}|U_{\alpha 4}|^2\lesssim
10^{-15}\left(\frac{1~\mathrm{TeV}}{m_4}\right)\left(\frac{g_*}{100}\right)^{1/2}, 
\end{equation}
with $g_*$ the effective degrees of freedom of relativistic particles.
One can verify that the order of magnitude of the mixing matrix
$|U_{\alpha 4}|$ required to generate
the baryon asymmetry of the Universe is much smaller than what 
would be required to obtain a large electron EDM (see Fig.~\ref{fig:edm2}). 
If the mixing matrix is fixed as large value, as $|U_{\alpha 4}|^2\sim10^{-3}$
as to obtain a large electron EDM, only a small baryon asymmetry would be
generated due to the strong washout effect.

\section{Summary}

We have computed the charged lepton EDMs at two-loop level in the
effective model and in the minimal Inverse Seesaw model. 
We found that at least two sterile fermions are needed to obtain a sizeable
electron EDM, within sensitivity reach of the upgraded ACME future experiment. 
In Inverse Seesaw models, the Dirac type contribution is dominant since
the pairs of heavy sterile fermions construct pseudo-Dirac fermions. 
In this case, the maximum value of the electron EDM is slightly below
the future sensitivity. 
The heavy sterile fermion masses should be in the range of
$100~\mathrm{GeV}-10~\mathrm{TeV}$ to induce large charged lepton EDMs
due to experimental and theoretical constraints.

\section*{Acknowledgments}
A. A. acknowledges support from EU H2020 grants No 690575 and No 674896.
T. T. acknowledges support from P2IO Excellence Laboratory (LABEX).

\end{document}